# Data-driven Hi2Lo for Coarse-grid System Thermal Hydraulic Modeling

**Arsen S. Iskhakov, Nam T. Dinh**
Department of Nuclear Engineering, North Carolina State University
Campus Box 7909, Raleigh, NC, USA 27695-7909
aiskhak@ncsu.edu; ntdinh@ncsu.edu

**Victor Coppo Leite, Elia Merzari**
Department of Nuclear Engineering, Pennsylvania State University
206 Hallowell Building, University Park, PA, USA 16802-4400
vbc5085@psu.edu, ebm5351@psu.edu

**ABSTRACT**

Traditional 1D system thermal hydraulic analysis has been widely applied in nuclear industry for licensing purposes due to its numerical efficiency. However, such codes are inherently deficient for modeling of multiscale multidimensional flows. For such scenarios coarse-grid 3D simulations are useful due to the balance between the cost and the amount of information a modeler can extract from the results. At the same time, coarse grids do not allow to accurately resolve and capture turbulent mixing in reactor enclosures, while existing turbulence models (closures for the Reynolds stresses or turbulent viscosity) have large model form uncertainties. Thus, there is an interest in the improvement of such solvers.
In this work two data-driven high-to-low methodologies to reduce mesh and model-induced errors are explored using a case study based on the Texas A&M upper plenum of high-temperature gas-cooled reactor facility. The first approach relies on the usage of turbulence closure for eddy viscosity with higher resolution/fidelity in a coarse grid solver. The second methodology employs artificial neural networks to map low-fidelity input features with errors in quantities of interest (velocity fields). Both methods have shown potential to improve the coarse grid simulation results by using data with higher fidelity (Reynolds-averaged Navier-Stokes and large eddy simulations), though, influence of the mesh-induced (discretization) error is quite complex and requires further investigations.

## 1. INTRODUCTION

System thermal hydraulic analysis historically relied on 1D codes due to their numerical efficiency. However, the results predicted by such codes are affected by errors due to closure model deficiencies, approximations in the numerical solution, nodalization effects, and imperfect knowledge of boundary and initial conditions, *etc.* [1]. To model multidimensional behavior such as mixing and thermal stratification in reactor enclosures, coarse-grid (CG) codes can be employed, *e.g.*, GOTHIC or SAM [2, 3]. However, CG (often not within the asymptotic converge region) does not allow to resolve multiscale turbulence effects in the flows, while turbulence models yield large uncertainties in the results [4]. One the other hand, despite the tremendous growth of computational power, direct resolution of fine-scale physics is infeasible for the licensing purposes. Therefore, there are continuous efforts aimed at improvement of the predictive capability of the system codes [5]. One of the most popular ways to do so is high-to-low (Hi2Lo) methodology, when a relatively high-fidelity tool (or data) is used to inform or calibrate a code with lower fidelity [6]. With recent development of data-driven (DD) approaches, machine learning (ML) is becoming popular for Hi2Lo and multiscale bridging [7]. There are numerous examples of using ML in fluid dynamics and thermal hydraulics for Reynolds-averaged Navier-Stokes (RANS) and other closures, *e.g.*, [7, 8, 9, 10]. However, a very limited number of papers addresses the challenges associated with usage of CG in simulations. For example, Bao et al. [11, 12, 13, 14] developed optimal mesh/model information system that allows choosing of optimal mesh size and closures for GOTHIC. A feedforward neural network (NN) was trained to predict error in low-fidelity quantities of interest (QoI), such as temperature, velocity, void fraction, *etc*. The error correction of low-fidelity simulation results is introduced *a posteriori*. On the contrary, Liu et al. [15] developed turbulence closure for SAM code using high-fidelity data, which allowed to consider transients with improved *in situ* turbulence modeling (eddy viscosity field was predicted by a convolutional NN).

Recently established IRP-NEAMS-1.1 Thermal-Fluids Applications in Nuclear Energy project seeks to deliver improved, fast-running models for complex physical phenomena involving turbulent mixing, thermal stratification, and thermal striping in complex geometries relevant to these reactors by addressing several challenge problems. Particularly, one of the objectives of challenge problem 3 [16] is to develop a ML-based framework for Hi2Lo informing and multiscale bridging of CG simulations using high-fidelity / high-resolution data. As the actual data are to be generated [17, 18], this paper uses a simplified case study to explore approaches developed by Bao et al. [11, 12, 13, 14] and Liu et al. [15] in application to jet injection and mixing in the upper plenum of 1/16$^{th}$ scaled high temperature gas-cooled reactor (HTGR) facility (Texas A&M University (TAMU)).

The rest of the paper is organized as follows. Section 2 provides details on the models and data generation activities; Section 3 explores Hi2Lo methodology that uses "high-resolution / high-fidelity" closure for eddy viscosity to inform CG simulations; Section 4 explores direct correction of errors in QoI (velocity). Section 5 concludes the paper and discusses the future work.

## 2. MATHEMETICAL MODELS AND DATA GENERATION

### 2.1. Case Study Description

Geometry of upper plenum of 1/16$^{th}$ scaled HTGR TAMU facility [19] is modeled, Figure 1. The upper plenum is represented by the slightly flattened semi-spherical volume with several (up to 25) inlet tubes at the bottom for injection of water jets. A single jet coming through the central tube with Reynolds number $Re_j = \rho V_j d_j / \mu = 12819$ case is considered. Water outflows from the annular periphery of the volume. In the experiments [19] characteristics of the jet at different elevations $H$ from the bottom are measured and mixing in the volume is observed. In this work dimensional (MOOSE) and non-dimensional (Nek5000/NekRS) simulations are performed, and fluid density and viscosity are fixed: $\rho = 10^3$ kg/m$^3$ and $\mu = 10^{-3}$ Pa·s, while jet inlet average velocity (vertical component) and diameter are adopted according to the experiments $V_j = 0.673$ m/s, $d_j = 19.05 \cdot 10^{-3}$ m.

### 2.2. Mathematical Models

The HTGR TAMU experiment [19] is modeled as isothermal and steady state. For simplicity, 2D axisymmetric numerical model is considered in CG and RANS simulations, while large eddy simulation (LES) is performed for 3D geometry.

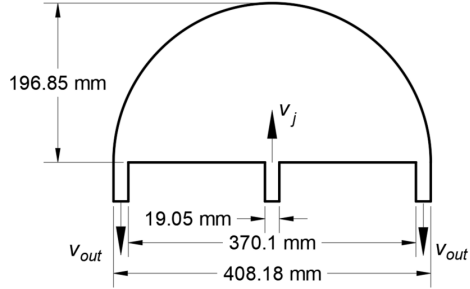

Figure 1. Sketch of the upper plenum of scaled HTGR TAMU facility.

Thus, the incompressible RANS equations with employed linear eddy viscosity hypothesis are being solved for CG and RANS simulations:

$$\nabla \cdot \mathbf{U} = 0$$
$$\rho \frac{D\mathbf{U}}{Dt} = -\nabla P + \nabla \cdot [\mu_{eff}(\nabla \mathbf{U} + (\nabla \mathbf{U})^T)] \quad (2.1)$$

where $\mathbf{U}(r,z) = \{U, V\}$ is Reynolds-averaged velocity, $D/Dt$ is substantial derivative, $P$ is modified pressure (by turbulent kinetic energy (TKE) component), $\mu_{eff} = \mu + \mu_t$ is effective viscosity, $\mu_t$ is eddy viscosity.

Finite volume (FV) incompressible MOOSE Navier-Stokes (NS) kernels [20] and spectral element method (SEM) solvers Nek5000/NekRS [21, 22] are used to perform the simulations. MOOSE has traditionally been a finite element (FE) framework [23], however, a poor performance (convergence issues) of FE kernels is observed for structured grids with sharp transition regions (which is almost inevitable). The motivation of using the structured grids is conditioned by the possibility to have a good control over the mesh quality. The MOOSE kernels utilize mixing length turbulence model [24]:

$$\mu_t = \rho l_m^2 (2S_{ij}S_{ij})^{1/2} \quad (2.2)$$

where $S_{ij} = 0.5(\partial U_i/\partial x_j + \partial U_j/\partial x_i)$ is mean rate of strain tensor, $l_m = 0.41 y_d$ is mixing length, $y_d$ is distance to wall.

RANS simulation using Nek5000 is performed using *k-τ* turbulence model [25]:

$$\mu_t = \rho C_\mu f_m k \tau \quad (2.3)$$

where $C_\mu = 0.09$, $f_m$ is wall damping function, $k$ is TKE, $\tau = k/\varepsilon$ is turbulent time scale, $\varepsilon$ is turbulent dissipation rate.

NekRS (a GPU-oriented version of Nek5000) is used to perform LES to obtain highly resolved data, which are verified against the experimental data in [17]. The LES solves filtered NS equations:

$$\nabla \cdot \bar{\mathbf{u}} = 0$$
$$\rho \frac{D\bar{\mathbf{u}}}{Dt} = -\nabla \bar{p} + \mu \nabla^2 \bar{\mathbf{u}} - \nabla \cdot \tau_{ij} \quad (2.4)$$

where overbar denotes filtered component, $\tau_{ij}$ is stress term closed by high-pass filter (HPF) model [26]:

$$\tau_{ij} - \frac{\delta_{ij}}{3}\tau_{kk} \approx -2(C_s\overline{\Delta})^2|\bar{s}_{ij}(H*\bar{\mathbf{u}})|\bar{s}_{ij}(H*\bar{\mathbf{u}}) \tag{2.5}$$

where $\overline{\Delta}$ is grid filter width, $C_s$ is model coefficient, $H*\bar{\mathbf{u}}$ denotes HPF quantity. The LES results are inherently time-dependent and were time-averaged and spatially collapsed into the *RZ* plane.

The boundary conditions are no-slip walls, free outflow at outlet, and fully developed turbulent profile at the inlet $V_{in} = 0.5V_j(1/7+1)(1/7+2)(1-2r/d_j)^{1/7}$ [27] (inlet tube has length $15d_j$ to reduce the bias in the specified profile).

### 2.3. Data Generation

Therefore, the following datasets are generated: (1) **CG datasets** that are obtained with the FV MOOSE kernels. Four meshes with different refinement factors (1x, 2x, 4x, 8x) are used to perform CG simulations, see Figure 2 (a-d); (2) **high-resolution (HiRe) dataset** that is also obtained with the FV MOOSE kernels, but uses the most refined mesh (16x), see Figure 2 (e); (3) **RANS dataset** that is referred to the results of Nek5000 simulations with *k-τ* model; (4) **LES dataset** that is referred to the LES results with the HPF model. RANS mesh has ~$10^6$ degrees of freedom (DoF). LES mesh has ~$5\cdot10^7$ DoF (in 3D).

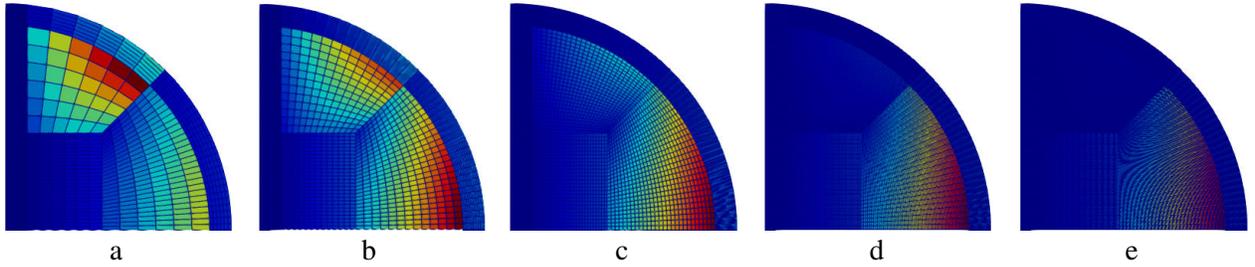

**Figure 2. Meshes used for FV MOOSE simulations. CG datasets: (a) coarsest (1x), 1494 elements; (b) 2x refined, 4147 elements; (c) 4x refined, 12393 elements; (d) 8x refined, 40645 elements. HiRe dataset: (e) 16x refined, 144189 elements. Color map shows the volume of mesh elements.**

Two major sources of uncertainties in CG and HiRe simulations can be outlined: (1) turbulence model form error that comes due to the eddy viscosity assumption and its formulation; (2) discretization error which is formally defined as the difference between the exact solution to the mathematical model and the exact solution to the discrete equations [28]. FV MOOSE kernels use 1st order upwind scheme for the advection term. Therefore, numerical viscosity plays a dominant role in the discretization error. Moreover, we recognize that some of the CG simulations might not be within the asymptotic mesh convergence region, which does not even guarantee the formal order of accuracy and makes the discretization error unpredictable. Quality of mesh might also play an important role, but in this work, it is assumed to have a small influence – future studies may include comparison of different strategies for the mesh blocking. RANS dataset also features these errors, but they are assumed to be lower than in CG and HiRe datasets. Finally, LES dataset also has the same sources of uncertainty, which are assumed to be lower than in RANS. Additionally, LES has a sampling error due to the finite temporal statistics, see Figure 3.

Figure 4 shows the comparison of velocity fields for CG (1x), HiRe, RANS, and LES data. One can clearly see that the jet velocity is overpredicted, while mixing in the volume is underpredicted comparing to the RANS and LES data. At the same time no significant difference is observed between CG and HiRe data, which suggests that the turbulence model error is much larger than the discretization error.

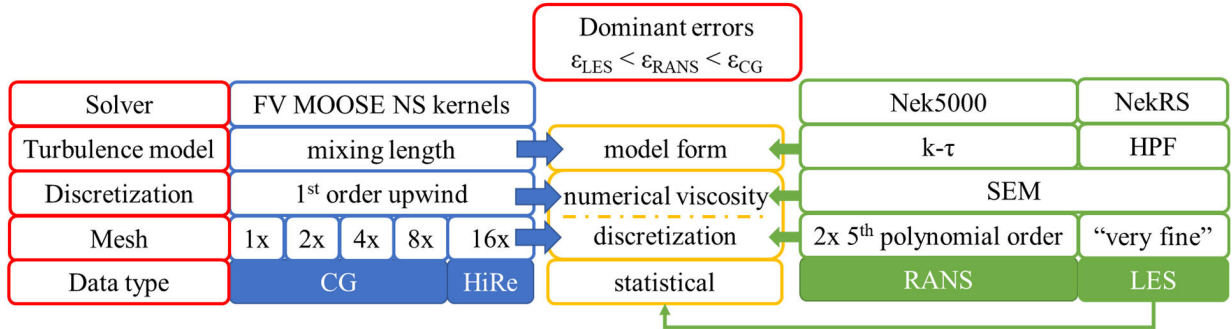

Figure 3. Generated datasets and major sources of uncertainties.

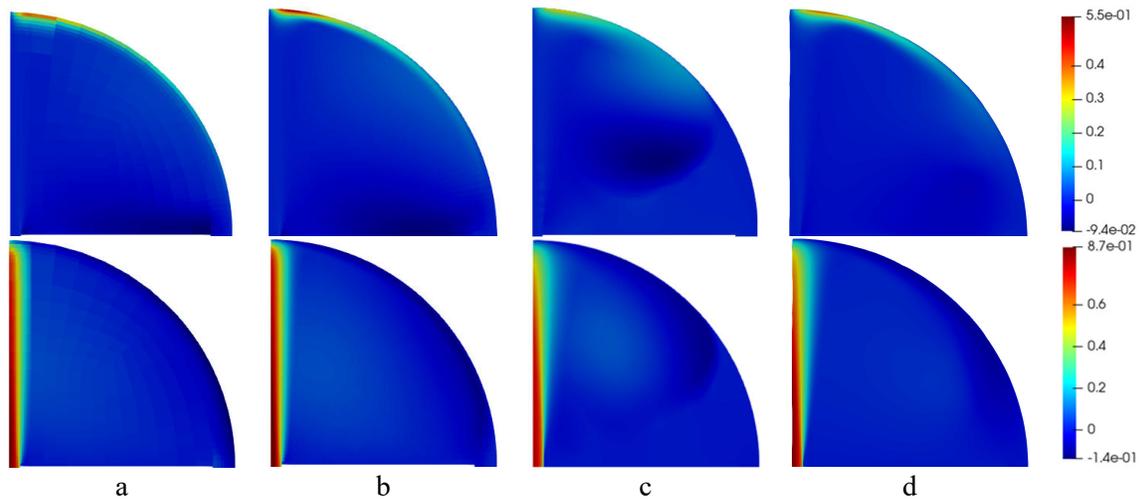

| a | b | c | d |

Figure 4. Horizontal (top) and vertical (bottom) velocity fields (m/s):
(a) CG (1x); (b) HiRe; (c) RANS, (d) LES.

## 3. HI2LO USING EDDY VISCOSITY CLOSURE

### 3.1. Hi2Lo using HiRe Eddy Viscosity

In this study HiRe eddy viscosity field $\mu_{t\_HiRe}$ is used in simulations on coarser grids (1x, 2x, 4x, and 8x). The objective is to test the possibility to reduce the discretization error by using a closure obtained within the same solver, but with higher resolution. This intrinsically assumes that the discretization error can be partitioned between the closure term ($\mu_t$) and QoI (**U**) and, by decreasing the first, one can decrease the total error. To implement the approach, $\mu_{t\_HiRe}$ field is saved to a .csv file. To allow its usage, a new FV MOOSE kernel is developed[1]. The kernel reads a pre-calculated eddy viscosity field from the file (only once, which is applicable to steady state simulations) instead of calculating it using the mixing length model. The point-to-point mapping is employed; alternatively, one can use cell-to-cell mapping that employs some local averaging [11], see Figure 5.

---

[1] https://github.com/aiskhak/moose_read_eddy_visc_from_file

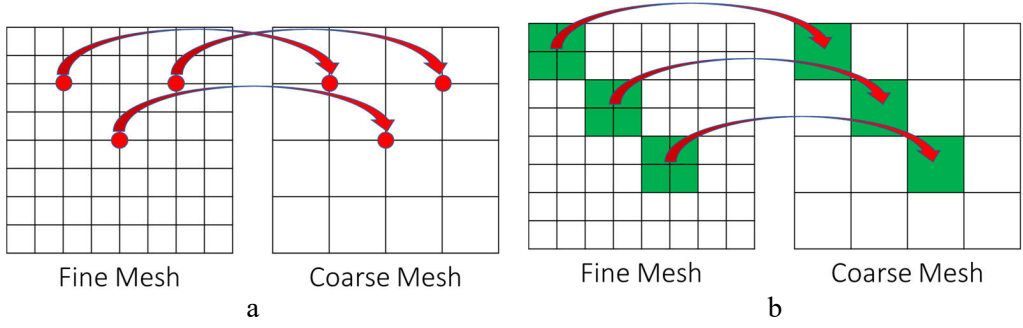

**Figure 5. (a) Point-to-point vs. (b) cell-to-cell mapping.**

First, CG velocity fields are compared against HiRe data. Figure 6 shows how the discretization error, $\varepsilon = QoI_{HiRe} - QoI_{CG}$, decreases as the mesh refinement is performed. Please note that the discretization error is computed using HiRe results since the exact solution to the mathematical model is unknown and the Richardson extrapolation [26] might not yield meaningful results for the 1st order scheme [28]).

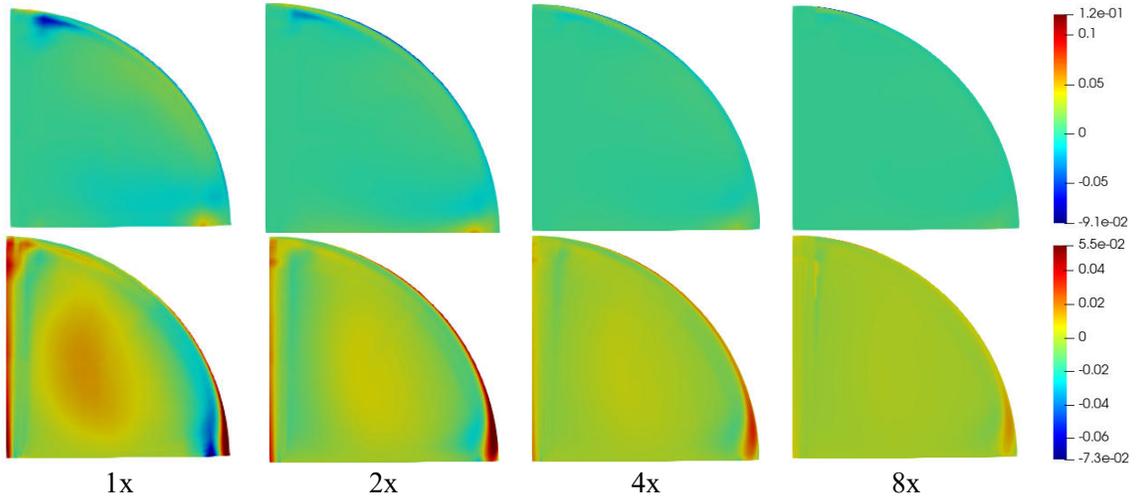

**Figure 6. Error in velocity fields (m/s) in CG simulations (comparing to HiRe data).**

Second, simulations are performed with $\mu_{t\_HiRe}$ and compared against HiRe data, $\varepsilon = QoI_{HiRe} - QoI_{CG}(\mu_{t\_HiRe})$, see Figure 7. Additionally, the following figure of merit (FoM) is calculated to compare the overall error metric based on the mean squared error (MSE) in the results (see Figure 8):

$$FoM = MSE(U_{HiRe} - U_{CG}) + MSE(V_{HiRe} - V_{CG}) \qquad (3.1)$$

Despite the expectations, usage of $\mu_{t\_HiRe}$ on coarser grids increases the overall error, which implies that the discretization error in $\mu_t$ does not play an important role in the overall error. One of the potential problems is introduction of non-smoothness in $\mu_t$ due to the point-to-point mapping, which can be reduced by using ML algorithms for closures developing, which is a matter of future work. However, even with smoother $\mu_{t\_HiRe}$, there is a little hope that the discretization error can be significantly reduced through the usage of HiRe closure from the same model and solver. Thus, the discretization error should be decreased by working directly with QoI, which is demonstrated in Section 4.

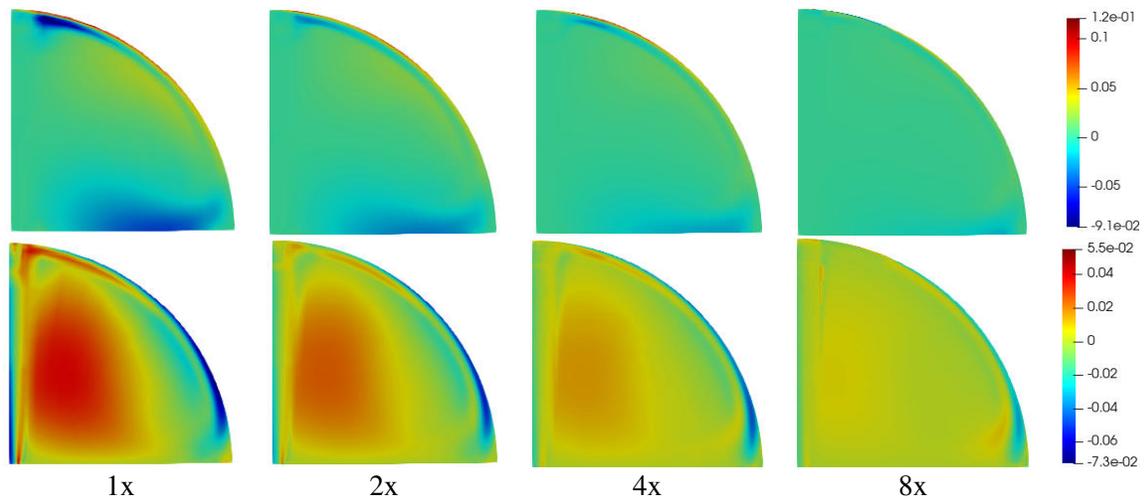
**Figure 7. Error in velocity fields (m/s) in CG simulations with $\mu_{t\_HiRe}$ (comparing to HiRe data).**

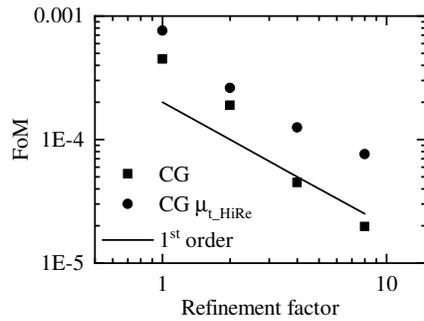
**Figure 8. FoM for CG and CG simulations with $\mu_{t\_HiRe}$ (comparing to HiRe data).**

### 3.2. Hi2Lo using RANS Eddy Viscosity

Comparing to the previous study, RANS eddy viscosity $\mu_{t\_RANS}$ is employed. The objective is to investigate the possibility to reduce the model form error in CG/HiRe simulations through a closure with higher fidelity as well as to study the influence of the discretization error. RANS and HiRe eddy viscosity fields are compared in Figure 9 (a, b). One can see a significant qualitative and quantitative difference in the fields.

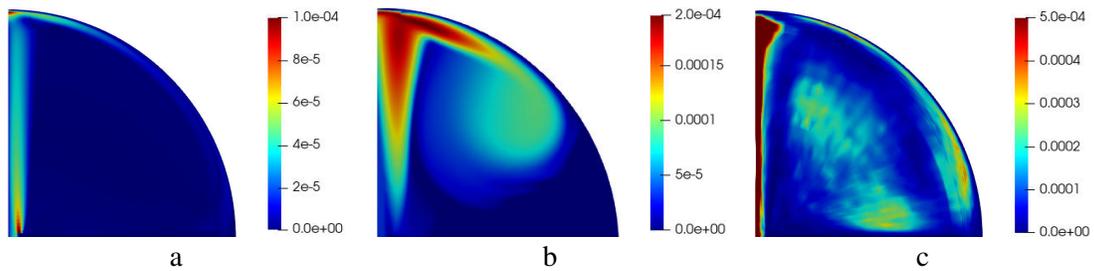
**Figure 9. (a) HiRe, (b) RANS, (c) LES eddy viscosity fields (Pa·s).**

Figure 10 shows the error in CG/HiRe simulations comparing to RANS data ($\varepsilon = QoI_{RANS} - QoI_{CG/HiRe}$). There is a significant difference, which is mostly caused by the turbulence model form error. At the same time, there is a little influence of the mesh refinement factor on the results.

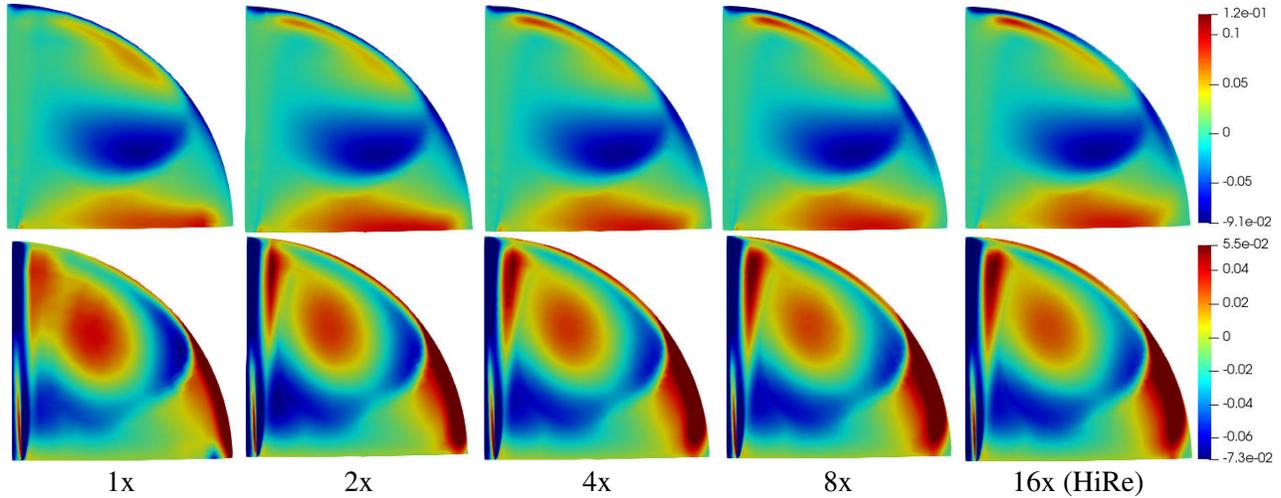

1x          2x          4x          8x          16x (HiRe)

**Figure 10. Error in velocity fields (m/s) in CG/HiRe simulations comparing to RANS data.**

Figure 11 shows the error for CG/HiRe simulations with RANS eddy viscosity, $\varepsilon = QoI_{HiFi} - QoI_{CG/HiRe}(\mu_{t\_RANS})$. One can see that in some regions (where the difference between the eddy viscosities is significant, Figure 9 (a, b)) the error is largely reduced. However, the remained regions still have large uncertainties. At the same time, there is no significant difference for different mesh resolutions: Figure 12 suggests that the discretization error plays ambiguous role in Hi2Lo results. Thus, the model form error can be significantly reduced by using a closure with higher fidelity, while the discretization error is unpredictable and must be separately addressed.

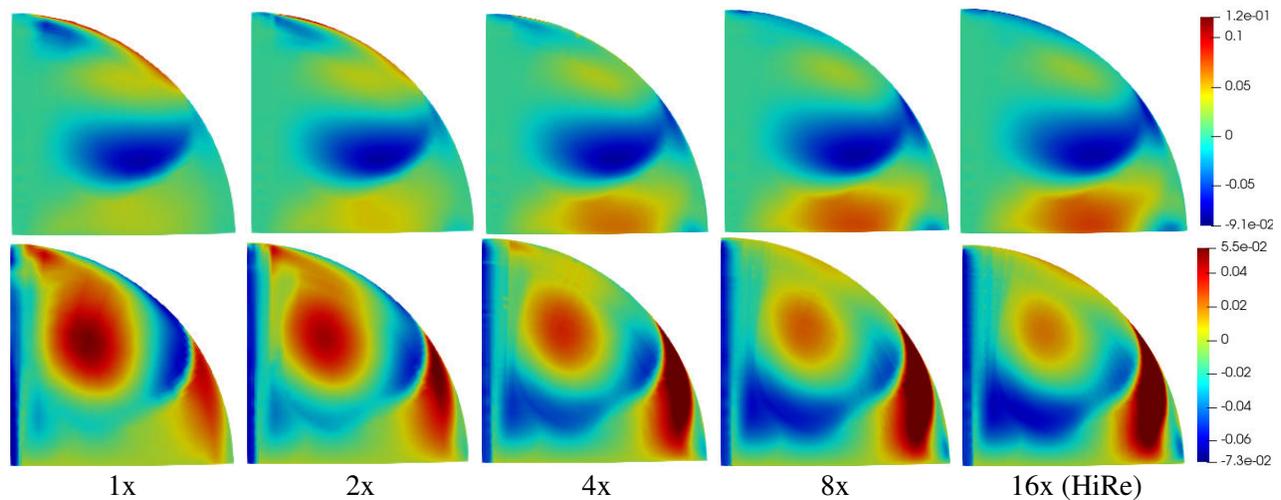

1x          2x          4x          8x          16x (HiRe)

**Figure 11. Error in velocity fields (m/s) in CG/HiRe simulations with $\mu_{t\_RANS}$ comparing to RANS.**

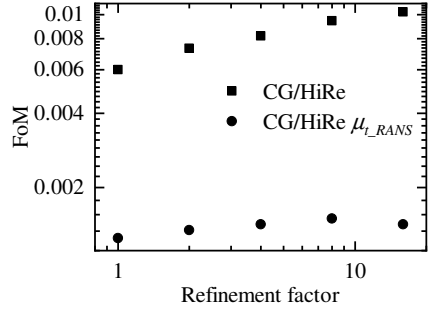

**Figure 12. FoM for CG/HiRe and CG/HiRe simulations with $\mu_{t\_RANS}$ (comparing to RANS data).**

### 3.3. Hi2Lo using LES Eddy Viscosity

In this study, eddy viscosity from time- and space-averaged LES is employed in order to test the applicability of the framework for the high-fidelity data. For eddy viscosity extraction, statistics for velocity derivatives $\partial U_i/\partial x_j$ and Reynolds stresses $<u'_i u'_j>$ is accumulated. Then, an approach based on the overall error minimization is adopted [29]:

$$\mu_t = \frac{\rho \sum_{j=1}^{3} \sum_{i=1}^{3} \left(\frac{2}{3} k \delta_{ij} S_{ij} - <u'_i u'_j> S_{ij}\right)}{\sum_{j=1}^{3} \sum_{i=1}^{3} (2 S_{ij} S_{ij})} \qquad (3.2)$$

where $\delta_{ij}$ is Kronecker delta. Given the deficiency of the eddy viscosity hypothesis [30] and finite temporal statistics accumulated, negative values of $\mu_{\mu_{t\_LES}t}$ are almost inevitable (nullified). After getting a 3D field of $\mu_{t\_LES}$ spatial averaging is performed. Note that alternatively one can do spatial averaging first (for $S_{ij}$ and $<u'_i u'_j>$), and then calculate $\mu_{t\_LES}$. This approach is more complex due to the tensor coordinate transformations involved (averaging of scalar $\mu_{t\_LES}$ does not require that). Figure 9 (c) shows $\mu_{t\_LES}$, which is noticeably different from HiRe and RANS; one can also notice some noisiness which comes from the limited statistics.

Extracted $\mu_{t\_LES}$ is used to replace the mixing length values in CG/HiRe simulations. Different behavior is observed – for coarse grids (1x, 2x, 4x) the results became worse, see Figure 13 (FoM is higher with $\mu_{t\_LES}$), while the overall error decreases as the mesh is refined. Such behavior can be caused by the following factors: (1) error in the extraction of $\mu_{t\_LES}$; (2) statistical error during the time averaging; (3) stronger influence of the mesh-induced error for high-fidelity data. These problems will be analyzed and addressed in the future work.

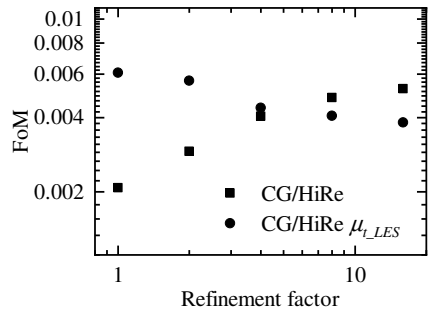

**Figure 13. FoM for CG/HiRe and CG/HiRe simulations with $\mu_{t\_LES}$ (comparing to LES data).**

## 4. HI2LO USING DIRECT ERROR CORRECTION IN QUANTITIES OF INTEREST

From the case studies in Section 3 one can outline at least two possible frameworks for using of ML for improvement of CG simulations: (1) *1-step approach*: reduce errors by regressing the discretization & model errors in QoI (no turbulence closure is needed). Below this approach is explored in application to the HTGR TAMU geometry. (2) *2-step approach*: (i) reduce turbulence model error by developing ML model using data with higher fidelity; (ii) reduce discretization error by correcting the QoI. In this case the total error is decomposed in order to increase the explainability of the results and (possibly) the efficiency. This is a more complex approach that will be explored in the future work.

### 4.1. Machine Learning Model Formulation and Training

To regress an unknown function that maps CG quantities (input features) with errors in QoI, a feedforward neural network (FNN) is developed. The hyperparameters of the FNN and other training details are presented in Table I; the FNN is implemented via Pytorch ML library [31]. Same procedure is performed for RANS and LES data to inform CG/HiRe results.

**Table I. FNN hyperparameters.**

| Parameter | Value |
|---|---|
| Input layer | 6 neurons |
| Hidden layers | 50-100-100-100-50-25 neurons |
| Output layer | 2 neurons |
| Activation | SELU and sigmoid in the output layer |
| Number of training epochs | 500 |
| Number of training datapoints | 161,904 |
| Number of validation datapoints | 28,571 |
| Number of test datapoints | 12,393 |
| Training batch size | 512 |
| Training algorithm | Adam |
| Loss function | MSE |
| Learning rate | $10^{-3}$ with decay on plateau |
| Dropout rate | 0.2 |

The following quantities are used as input features from the CG/HiRe simulations:
- velocity derivatives $\frac{\partial U}{\partial r}, \frac{\partial U}{\partial z}, \frac{\partial V}{\partial r}, \frac{\partial V}{\partial z}$ (normalized in [-1,1]);
- eddy viscosity $\mu_t$, assuming it is connected with the model form error ([0,1]);
- volume of mesh elements $V_{mesh}$, assuming it is connected with discretization error ([0,1]).

The targets are errors in velocities:
- $U_{er} = U_{RANS/LES} - U_{CG/HiRe}$ and $V_{er} = V_{RANS/LES} - V_{CG/HiRe}$ ([0,1]).

Four datasets are used for training: CG simulations (1x, 2x, 8x refined mesh and HiRe (16x) simulation). One dataset (CG with 4x refined mesh) is used as a test one (completely unseen by the FNN). T-distributed stochastic neighbor embedding (t-SNE) dimensionality reduction algorithm is used to visualize the similarity of the datasets, see Figure 14. One can see that all datasets share some similarity; the test dataset (4x) has the largest spreading in the 2D map, which potentially means that it is hardest one to predict.

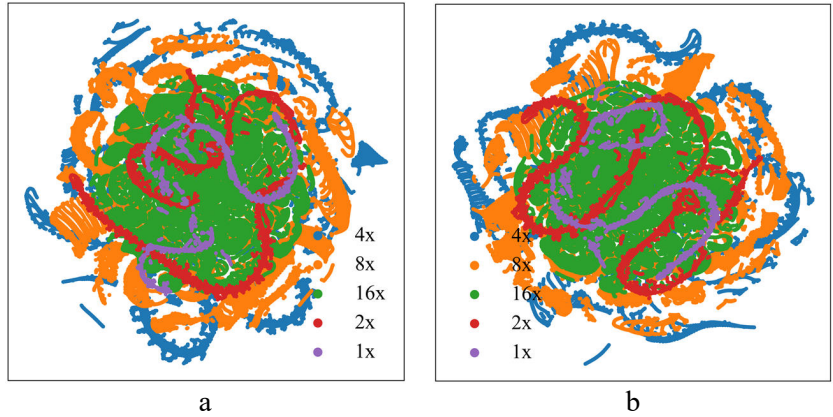

**Figure 14. t-SNE results for (a) RANS data; (b) LES data.**

### 4.2. Hi2Lo using RANS Data

After prediction of $\mathbf{U}_{er} = \mathbf{U}_{RANS} - \mathbf{U}_{CG/HiRe}$ for the test dataset, its velocity fields are corrected as $\mathbf{U}_{Hi2Lo} = \mathbf{U}_{CG/HiRe} + \mathbf{U}_{er}$. Figure 15 shows the improved fields. Figure 16 compares the corrected profiles at different elevations with CG, HiRe, and RANS data. One can see that the Hi2Lo results are almost identical to RANS data (compare Figure 15 to Figure 4 (c)) for both – jet velocity and mixing in the volume. Despite such drastic improvement one should keep in mind that with such point-wise mapping some discontinuities (non-smoothness) are introduced. Additionally, the conservation equations Eq. (2.1) are not obeyed anymore. Slightly more consistent results may be provided by convolutional NNs [15] by performing field-wise predictions. However, the major drawback is that such NNs are strictly linked to a particular geometry. Some physics-informed NN formulation penalizing the unphysical predictions can be potentially employed.

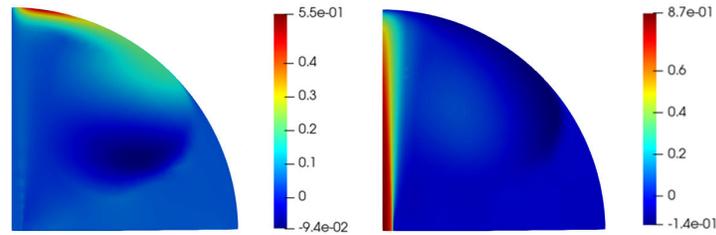

**Figure 15. RANS-Hi2Lo horizontal and vertical velocity fields (m/s).**

### 4.3. Hi2Lo using LES Data

Similarly to RANS-Hi2Lo, predicted error $\mathbf{U}_{er} = \mathbf{U}_{LES} - \mathbf{U}_{CG/HiRe}$ is used to correct CG/HiRe velocity fields, see Figure 17 Figure 18 compares the improved profiles with CG, HiRe, and LES data. The corrected results are very close to LES data; however, one can see noticeable "noisiness" in the corrected fields. Since there is no principal difference between the time-averaged velocity fields from LES and RANS, this can be caused by the noisiness in the LES data due to the statistical error.

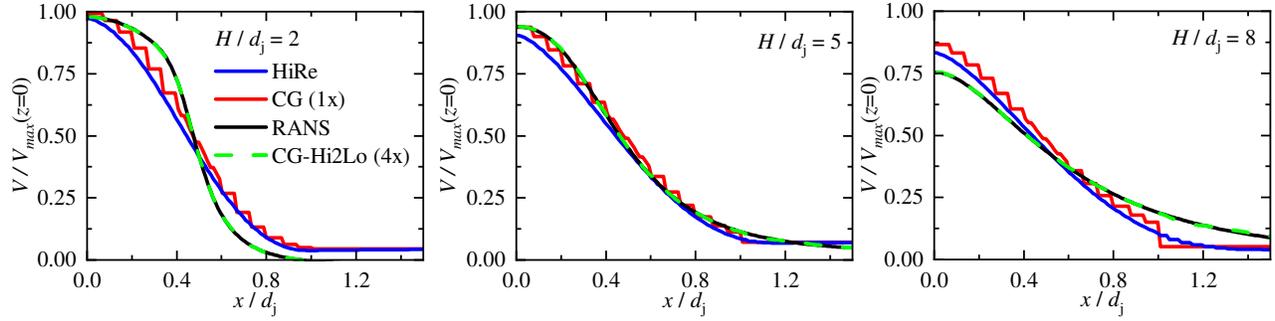

Figure 16. Comparison of jet vertical velocity profiles at elevation $H$ (Hi2Lo using RANS data).

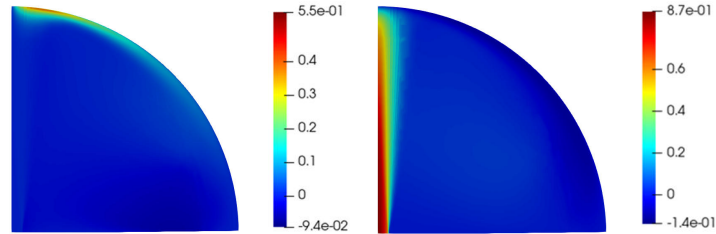

Figure 17. LES-Hi2Lo horizontal and vertical velocity fields (m/s).

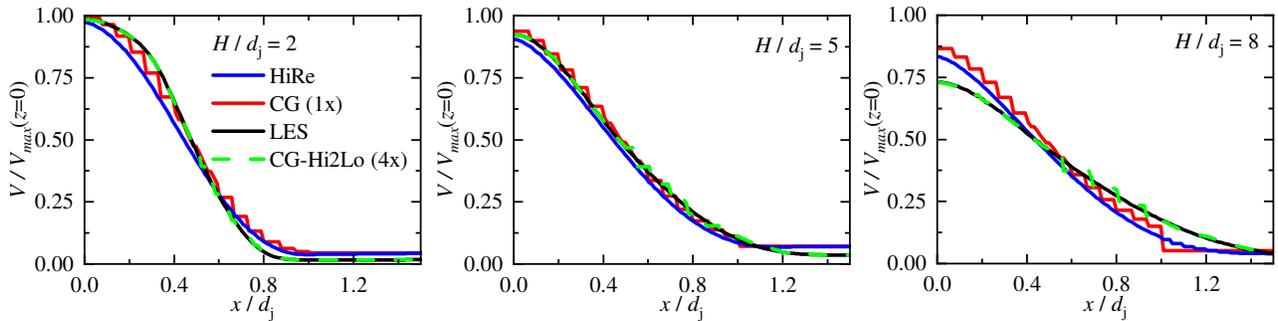

Figure 18. Comparison of jet vertical velocity profiles at elevation $H$ (Hi2Lo using LES data).

## 5. CONCLUSIONS AND FUTURE WORK

This paper investigated two Hi2Lo methodologies for improvement of CG simulation results. As a CG solver, incompressible FV MOOSE NS kernels with mixing length turbulence model are employed. Data with higher fidelity are generated using Nek5000/NekRS (RANS with $k$-$\tau$ turbulence model and LES with HPF model). It is demonstrated that turbulence model form errors can be largely decreased by using eddy viscosity with higher fidelity (RANS data). At the same time, extraction of eddy viscosity from LES data and noisiness due to limited statistics can bring large uncertainties. Furthermore, mesh (discretization) error has complex behavior and cannot be effectively reduced through the closure. In the second approach, a direct correction of errors in velocity is implemented. FNNs are trained to predict errors in CG velocity fields by using information from RANS and time averaged LES. It is shown that the total error can also be noticeably decreased with that methodology. Again, noisiness of the time averaged LES data creates additional difficulties in the process comparing to RANS-Hi2Lo. The major drawback of the approach is that the corrected quantities do not follow the conservation laws and its applicability to transients is questionable due to the possible convergence issues. The impossibility to obey the conservation laws also makes the corrected fields non-smooth.

Future work will be aimed at further investigation of the developed approaches including development of ML-based turbulence closures using high-fidelity data, separation of discretization and model form errors (2-step framework), consideration of different experimental conditions, modeling of transients, and studying of the influence of the discretization error under different mesh blocking strategies.

**ACKNOLEDGEMENTS**

This work was funded by a U.S. Department of Energy Integrated Research Project entitled "Center of Excellence for Thermal-Fluids Applications in Nuclear Energy: Establishing the knowledgebase for thermal-hydraulic multiscale simulation to accelerate the deployment of advanced reactors – IRP-NEAMS-1.1: Thermal-Fluids Applications in Nuclear Energy". The authors are thankful for all IRP Challenge Problem 3 participants for useful comments and recommendations.